\documentclass[11pt,twoside]{article}
\usepackage{./asp2010}

\resetcounters

\begin{document}

\title{X-ray Modeling of $\eta$ Carinae and WR140 from SPH Simulations}
\author{Russell, C. M. P.$^1$, Owocki, S. P.$^1$, Corcoran, M. F.$^{2,3}$, Okazaki, A. T.$^4$, and Madura, T. I.$^{1,5}$
\affil{$^1$Bartol Research Institute, Department of Physics and Astronomy, University of Delaware, Newark, DE, 19716, USA}
\affil{$^2$CRESST and X-ray Astrophysics Laboratory, NASA/GSFC, Greenbelt, MD, 20771, USA}
\affil{$^3$Universities Space Research Association, 10211 Wincopin Circle, Suite 500, Columbia, MD, 21044, USA}
\affil{$^4$Faculty of Engineering, Hokkai-Gakuen University, Toyohira-ku, Sapporo 062-8605, Japan}
\affil{$^5$Max-Planck-Institut f\"ur Radioastronomie, Auf dem H\"ugel 69, D-53121 Bonn, Germany}}

\begin{abstract}
The colliding wind binary (CWB) systems $\eta$ Carinae and WR140 provide unique laboratories for X-ray astrophysics. Their wind-wind collisions produce hard X-rays that have been monitored extensively by several X-ray telescopes, including RXTE. To interpret these X-ray light curves and spectra, we apply 3D hydrodynamic simulations of the wind-wind collision using smoothed particle hydrodynamics (SPH), with the recent improvements of radiative cooling and the acceleration of the stellar winds according to a $\beta$ law. For both systems, the 2-10 keV RXTE light curves are well-reproduced in \emph{absolute} units for most phases, but the light curve dips associated with the periastron passages are not well matched. In WR140, the dip is too weak, and in $\eta$ Carinae, the large difference in wind speeds of the two stars leads to a hot, post-periastron bubble that produces excess emission toward the end of the X-ray minimum.
\end{abstract}

\section{SPH Models \& X-ray Radiative Transfer}

To model the long-period, highly eccentric colliding wind binary (CWB) systems $\eta$ Carinae and WR140, we present 3D smoothed particle hydrodynamics (SPH) simulations of their wind-wind interaction (see Fig.~1), using the same general parameters as \citet{RussellP11b}.  As improvements upon the previous work of \citet{OkazakiP08} and \citet{RussellP11a, RussellP11b}, the SPH particles are accelerated (using an anti-gravity approach) to mimic a $\beta$\,=1 velocity law, and radiative cooling is explicitly treated with the exact integration scheme of \citet{Townsend09}.

The SPH visualization program \textsc{splash} \citep{Price07} has been modified to calculate the absolute flux of the model X-rays from ray-tracing of the emission $j=n_e n_H \Lambda(E,T)$ in the SPH simulations (see Fig.~1), with the emissivity $\Lambda(E,T)$ computed from the APEC plasma code \citep{APEC}.  Separate energy-dependent opacities, computed from \textit{windtabs} \citep{LeuteneggerP10}, are used for the O star and WC star in WR140.

\begin{figure}[t]
    \includegraphics[width=5.25in]{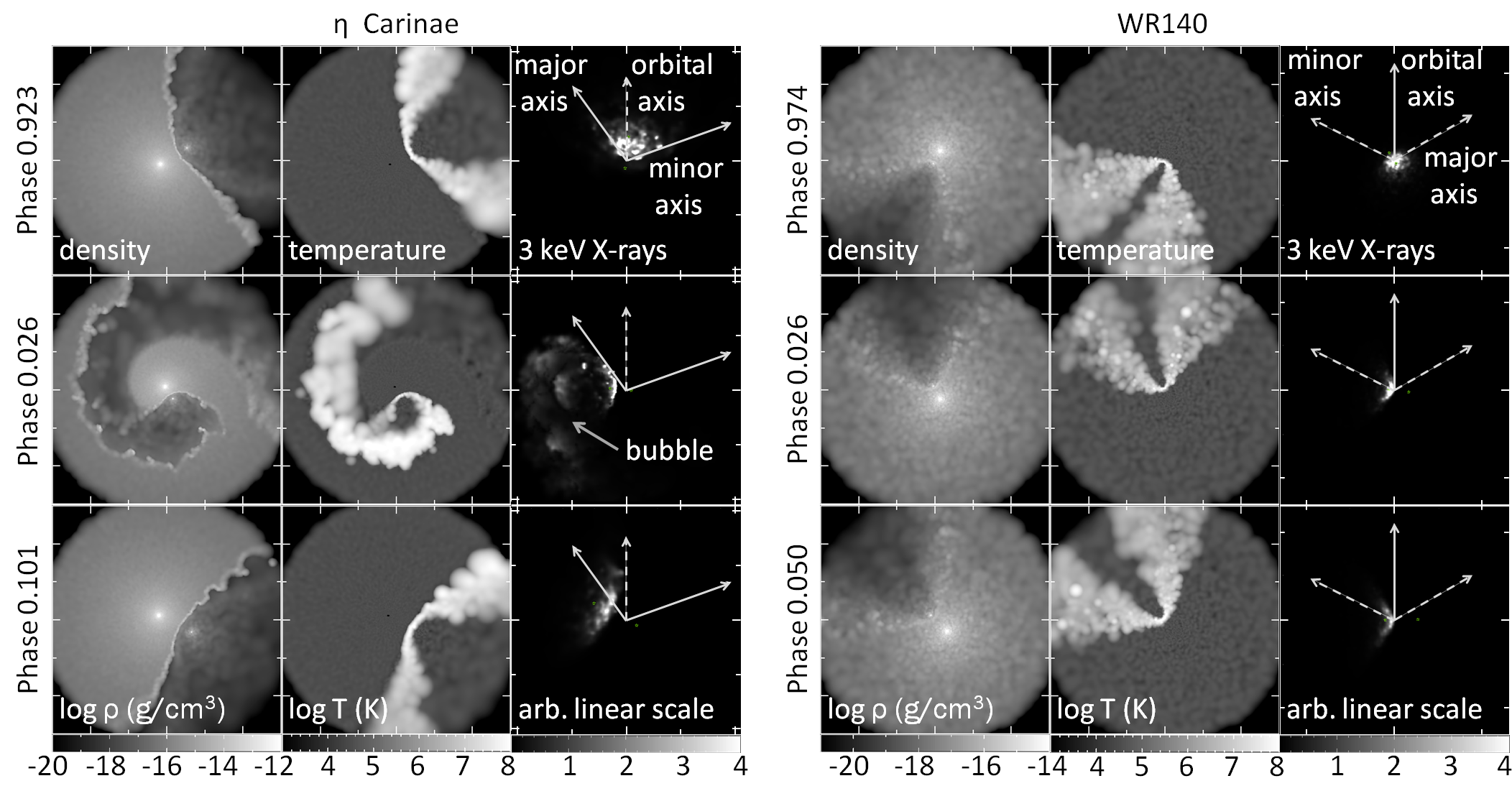}
 \caption{Density (left) and temperature (center) in the orbital plane of the SPH simulations, and 3 keV X-rays as observed from Earth (right). Each tick mark is $1a$. The X-ray emission is localized around the apex of the wind-wind interaction region, except when the post-periastron bubble occurs in $\eta$ Carinae.}
\end{figure}

\section{Results: Light Curves, Spectra \& Hardness Ratios}

Figure 2 compares the 2-10 keV RXTE light curve of $\eta$ Carinae \citep{CorcoranP10} and WR140 \citep{CorcoranP11} with our model results computed from \textsc{splash}\footnote{For technical reasons, the RXTE light curve is in flux for $\eta$ Carinae and counts for WR140. We take this into account by folding the WR140 model light curve through the RXTE response matrix.}. Both comparisons are in \emph{absolute} units, and given this, the level of agreement, apart from periastron, is quite remarkable, though there is some cycle-to-cycle variability not accounted for in the models.  Moreover, figure 3 shows that, for WR140, the model and observed apastron spectra also agree, while the model hardness ratio has a similar phase variation to that observed, but with an overall softer level that is likely due to the limited SPH resolution of the shock-heated material.

In contrast, near the periastron X-ray minima, both models overpredict the X-ray flux.  In $\eta$ Carinae, much of this excess results from the fast secondary wind catching up with the slow primary wind inducing a shock that blows a hot, extended, X-ray emitting, post-periastron bubble into the primary wind.  (Curiously, \citet{ParkinP11} find that this bubble does not produce much excess emission in X-rays.)  In WR140, the nearly equal wind speeds mean there is no such bubble, and so the more modest disagreement is just in the depth of the absorption minimum.

Overall, the WR140 model seems to need only minor tweaking, but in $\eta$ Carinae, there are some unknown, fundamentally important effects that are still not understood.

\begin{figure}[h]
    \includegraphics[width=5.25in]{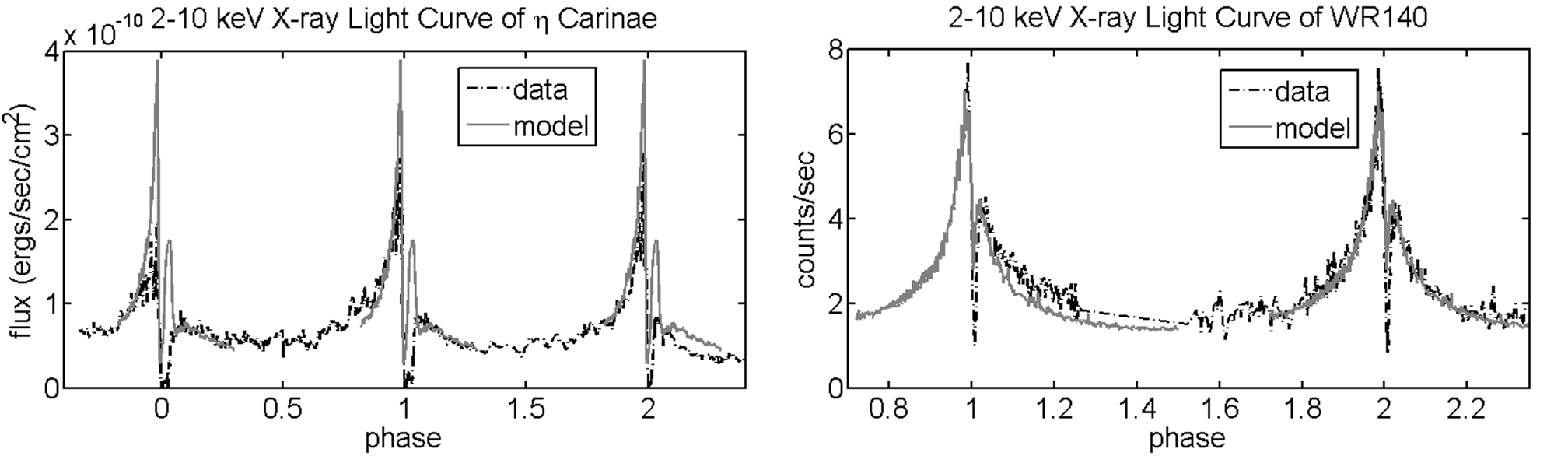}
 \caption{Model RXTE light curves plotted against the data in absolute units.  To account for the Homunculus' absorption in $\eta$ Carinae, an absorbing column of $N_H=5$e22 cm$^{-2}$ \citep{HamaguchiP07} was added.}
\end{figure}

\begin{figure}[h]
    \includegraphics[width=5.25in]{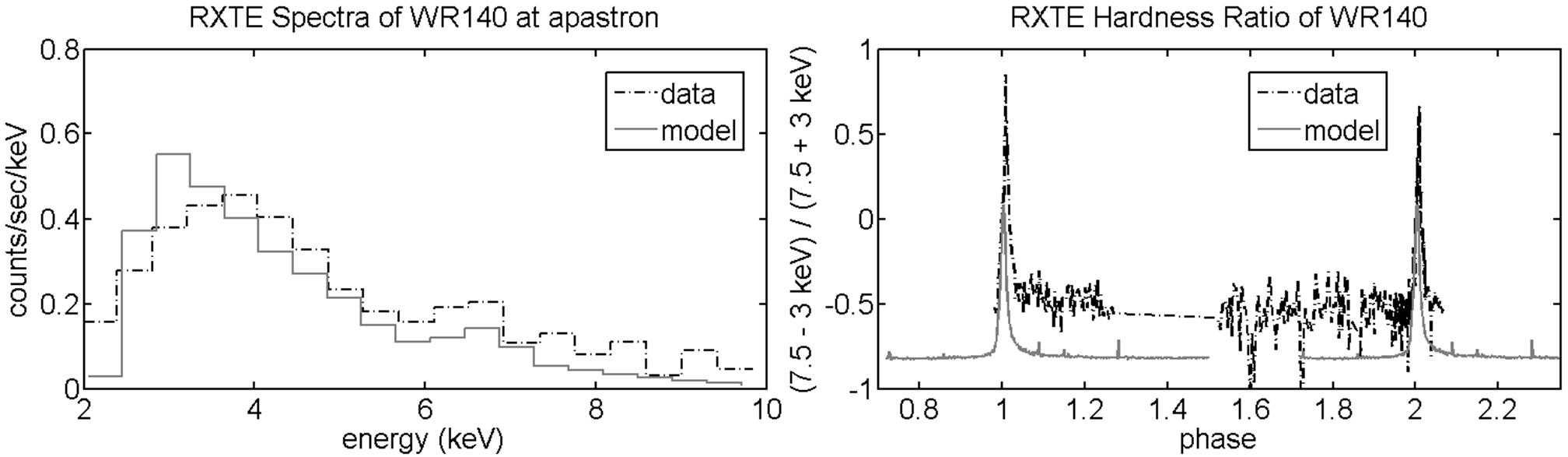}
 \caption{Model RXTE apastron spectra and hardness ratio of WR140, where the hardness ratio is the 7.5 keV flux minus the 3 keV flux divided by their sum.}
\end{figure}

\acknowledgements CMPR acknowledges support from the NASA Graduate Student Researchers Program \#NNX08AT36H and the NASA Delaware Space Grant Fellowship \#NNX10AN63H.  All SPH simulations were run on the Schirra and Pleiades supercomputers at the NASA Ames Research Center.

\bibliographystyle{asp2010}
\bibliography{Quebec2011Final}

\end{document}